# Universal Impedance Fluctuations in Wave Chaotic Systems

Sameer Hemmady,[1] Xing Zheng,[2] Edward Ott,[1,2] Thomas M. Antonsen,[1,2] and Steven M. Anlage[1,3]
Physics Department, University of Maryland, College Park, MD 20742-4111 USA

(7 May, 2004)

We experimentally investigate theoretical predictions of universal impedance fluctuations in wave chaotic systems using a microwave analog of a quantum chaotic infinite square well potential. Our approach emphases the use of the *radiation impedance* to remove the non-universal effects of the particular coupling from the outside world to the scatterer. Specific predictions that we test include the probability distribution functions (PDFs) of the real (related to the local density of states in disordered metals) and imaginary parts of the normalized cavity impedance, the equality of the variances of these PDFs, and the dependence of the universal PDFs on a single control parameter characterizing the level of loss. We find excellent agreement between the statistical data and theoretical predictions.

There is interest in the small wavelength behavior of quantum (wave) systems whose classical (ray orbit) limit is chaotic. Despite their apparent complexity, quantum chaotic systems have remarkable universal properties. Much prior work has focused on identifying the universal statistical properties of wave chaotic systems such as quantum dots and atomic nuclei [1-3]. For example, the nearest neighbor energy level spacing statistics of these systems have universal distributions that fall into one of three classes, depending on the existence or absence of time-reversal symmetry and symplectic properties. Likewise the eigenfunctions of wave chaotic systems have universal statistical properties, such as one-point and two-point statistical distribution functions [4-6]. It has been challenging to experimentally measure the corresponding universal properties of the scattering and impedance matrices of lossy multi-port wave chaotic systems. Here, we experimentally examine universal statistical properties of the complex impedance (or scattering) matrix fluctuations of such systems.

We consider wave systems in the semiclassical limit consisting of enclosures that show chaos in the ray limit, but which are also coupled to their surroundings through a finite number of leads or ports, and also include loss. Examples include quantum dots together with their leads, wave chaotic microwave or acoustical cavities together with their coupling ports, or scattering experiments on nuclei or atoms. Theoretical studies of chaotic scattering have examined the eigenphases of the scattering (S) matrix,[7] the time-delay distribution,[8,9,10] and the distribution of the scalar reflection and transmission coefficients.[10,11,12,13] Experimental work has concentrated on the energy decay and S-matrix autocorrelation functions in chaotic systems,[14,15,16,17] and most recently the distribution of scalar reflection coefficients[18]. In the related field of statistical electromagnetism [19], the statistical distribution of electromagnetic fields [20] and impedance [21] within complicated enclosed systems has been studied, but these results have not been generalized to other wave chaotic systems.

Ref. [18] (and references therein) takes into account both coupling and absorption in order to apply predictions of Random Matrix Theory (RMT) for the scattering matrices of real systems. There, the average of the reflection coefficient was measured in chaotic microwave cavities, and excellent agreement was found with the predictions of RMT. To remove the non-universal effect of the coupling configuration, Ref. [18], as advocated in previous theoretical approaches [22], uses the reflection coefficient averaged over a frequency range $\Delta f$ that is small compared to f but large compared to the mean mode frequency spacing. Thus the data for the chaotic cavity with losses is used to extract the universal properties from the same data. Since these non-universal properties are due to the detailed coupling geometry, it would seem useful to experimentally extract the characterization of the non-universal coupling from a measurement that depends only on the coupling geometry, and not on the cavity geometry and losses. In addition, it would also seem desirable to obtain this characterization from a single measurement at one frequency rather than as an average over frequency. This is achieved here for the first time through measurement of the radiation impedance of the ports.

Our purpose is to test specific predictions of RMT using an experiment that allows contact to both the quantum chaotic and wave chaotic aspects of the problem. We use a quasi-two-dimensional chaotic microwave resonator [23] (see insets in Fig. 1) to experimentally study the impedance and scattering matrices of wave chaotic cavities, including the coupling ports. Through the Helmholtz-Schrödinger analogy for two-dimensional electromagnetic cavities, our results also apply to quantum chaotic systems, such as quantum dots. We experimentally test the key theoretical predictions (which can be difficult to do in mesoscopic systems) for one-port systems and find very good agreement.

---

[1] Also with the Department of Electrical and Computer Engineering
[2] Also with the Institute for Research in Electronics and Applied Physics
[3] Also with the Center for Superconductivity Research

The theoretical predictions for the impedance matrix are universal and apply to all wave chaotic systems. For example, the real part of the impedance matrix of a wave chaotic system is directly related to the local density of states of mesoscopic metal particles [24]. The NMR lineshape of nm-scale metal clusters is predicted to be given by the PDF of Re[ $z$ ] given below [24,25], and is in good agreement with experiment [26]. Wigner introduced a related quantity, the R-matrix, as an alternative method to describe scattering problems [27] in quantum mechanics. In this case, space is divided into two parts, a finite interior domain containing the scattering potential of interest, and the remaining external asymptotic region. The wavefunction for the particle in the interior domain is expressed as a linear combination of discrete bound states, while the particle is described by scattering states in the external asymptotic region. The R-matrix describes the boundary condition linearly relating the normal (n) derivative of the wavefunction ($\hat{\psi}$) to the wavefunction itself, at the boundary of the interior domain as $\hat{\psi} = \vec{R}\ \partial\hat{\psi}/\partial n$. The analogous quantity in an electromagnetic system is the impedance matrix $\vec{Z}$, relating voltages ($\hat{V}$) and currents ($\hat{I} \sim \partial \hat{V}/\partial$n) at the ports as $\hat{V} = \vec{Z}\hat{I}$. The impedance matrix is related to the scattering matrix through $\vec{S} = \vec{Z}_0^{1/2}(\vec{Z}+\vec{Z}_0)^{-1}(\vec{Z}-\vec{Z}_0)\vec{Z}_0^{-1/2}$, where $\vec{Z}_0$ is a diagonal real matrix whose elements are the characteristic impedances of the transmission line modes connected to each port, and reduces to $Z = Z_0(1+S_{11})/(1-S_{11})$ for a one-port network.

Here we do not frequency average to extract universal fluctuation statistics. Rather, the primary new information required is knowledge of the radiation impedance for each port (or lead) of the system [28]. The radiation impedance depends only on the configuration of the coupling geometry and not on the cavity boundaries. It is experimentally accessible, system specific (hence nonuniversal), and independent of global properties of the system (e.g., chaos).

Reference [28] suggests a normalization procedure to remove the dependence on details and non-universal properties, and to reveal the underlying intrinsic fluctuations of the wave chaotic impedance. This involves the radiation impedance $Z_{Rad}$ that is the impedance seen at the input to the coupling structure for the same coupling geometry, but with the sidewalls of the cavity removed to infinity, so that no reflected waves come back to the port. The normalized impedance is, $z = \{\text{Re}[Z_{Cav}] + i(\text{Im}[Z_{Cav}] - \text{Im}[Z_{Rad}])\}/\text{Re}[Z_{rad}]$. In the case of a lossless cavity, the imaginary part is expected to exhibit a Lorentzian distribution with unit width [28, 32]. Losses manifest themselves through a finite real part of $z$ and also lead to the truncation of the tails of the lossless (Lorentzian) distribution of Im[ $z$ ]. It is also predicted that the variances of the Re[ $z$ ] and Im[ $z$ ] PDFs are equal and given by, $\sigma_{\text{Re}z}^2 = \sigma_{\text{Im}z}^2 = Q/(\pi \tilde{k}^2)$, in the case of time-reversal symmetric wave chaotic systems (Gaussian Orthogonal Ensemble). Here $\tilde{k}^2 = k^2/\Delta k^2$ where $k$ is the free space wavenumber, $\Delta k^2$ is the mean spacing in $k^2$ eigenvalues for the closed cavity ($\Delta k^2 = 4\pi/A$ for a two-dimensional cavity of area $A$), and $Q$ is the quality factor of the enclosure. The parameter $\tilde{k}^2/Q$ represents the ratio of the resonance width to the mean level spacing, similar to that defined in other RMT treatments [13,25]. The theory also makes quantitative predictions for the PDFs of Re[ $z$ ] and Im[ $z$ ] with this parameter.

For our experimental tests of the theory the driving port consists of the center conductor of a coaxial cable that extends from the top lid of the cavity and makes contact (shorts) with the bottom plate (Fig. 1(b) right inset), injecting current into the bottom plate of the cavity. This gives rise to a normalized resonance width due to coupling $\tilde{k}^2/Q_{coupling} \approx 0.03 - 0.12$ in this experiment between 7.2 and 8.4 GHz, depending on antenna and cavity geometry. This cavity has previously demonstrated a crossover from GOE to GUE statistics in the eigenvalue spacing statistics [29] and eigenfunction statistics [30,31]. To perform ensemble averaging, two perturbations (gray rectangles in fig 1(a) inset), made up of rectangular ferromagnetic solids wrapped in Al foil (dimensions 26.7 x 40.6 x 7.87 mm$^3$), are systematically scanned and rotated throughout the volume of the cavity by means of a strong magnet that is placed outside the cavity.

As predicted, the universal cavity impedance statistics can be drawn from $S_{11}$ measurements of the cavity (referred to as the Cavity Case) and a measurement with identical coupling, but with the walls of the cavity removed to infinity (the Radiation Case). The latter condition is realized experimentally by placing microwave absorber (ARC tech- DD10017D, <-25 dB return loss between 6 and 12 GHz) along all the sidewalls of the cavity. An ensemble of wave chaotic cavities is obtained by recording $S_{11}$ as a function of frequency (8001 points between 6 and 12 GHz) for 100 different positions and orientations of the perturbations within the cavity.

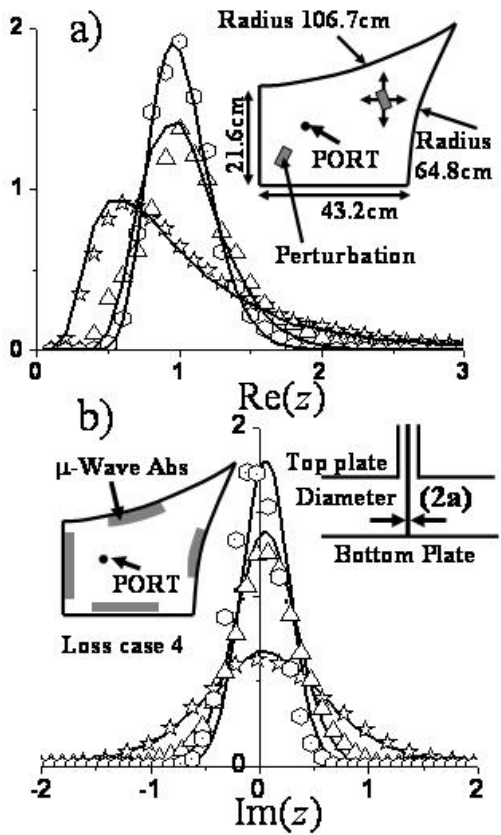

**Fig.1:**. PDFs for the (a) real and (b) imaginary parts of the normalized cavity impedance $z$ for a wave chaotic microwave cavity between 7.2 and 8.4 GHz with $h = 7.87$ mm and $2a = 1.27$ mm, for three values of loss in the cavity (open stars: 0, triangles: 2, hexagons: 4 strips of absorber). Also shown are single parameter simultaneous fits for both PDFs. The inset in (a) shows the cavity and the position of the coupling port. The inset on the left in (b) shows the realization of a high loss cavity (4 strips of absorber); while the inset on the right shows the details of the antenna in cross section.

Fig. 1 shows the evolution of the PDFs for the normalized cavity impedance in the presence of increasing loss. The losses are incrementally increased within the cavity by placing 15.2 cm-long strips of microwave absorber along the inner walls of the cavity. The inset in Fig. 2 shows the reflection spectra $|S_{11}|$ for the Radiation case as well as Loss Case 0 (which corresponds to the empty cavity with no microwave absorber strips) and Loss Case 4 (Fig. 1 left inset). The data shows that as the losses within the cavity increase, the PDF of the normalized imaginary part of the impedance loses its long tails and begins to sharpen up, developing a Gaussian appearance. The normalized PDF of the real part smoothly evolves from being peaked below 1, into a Gaussian-like distribution that peaks at 1 and sharpens with increasing loss. The PDF data is overlaid with a single-parameter fit to the theory, which simultaneously fits both the real and imaginary histograms for each loss scenario. There is a close overlap between the theoretical prediction and the experimental results with a choice of $\tilde{k}^2/Q = 0.8$, 4.2 and 7.6 in order of increasing loss (i.e. imaginary part of the resonant frequency). This is in good agreement with the typical $Q$ values for the cavity (e.g. about 200 at $\tilde{k}^2/Q = 0.8$) extracted from $S_{11}(\omega)$ measurements for these different loss scenarios.

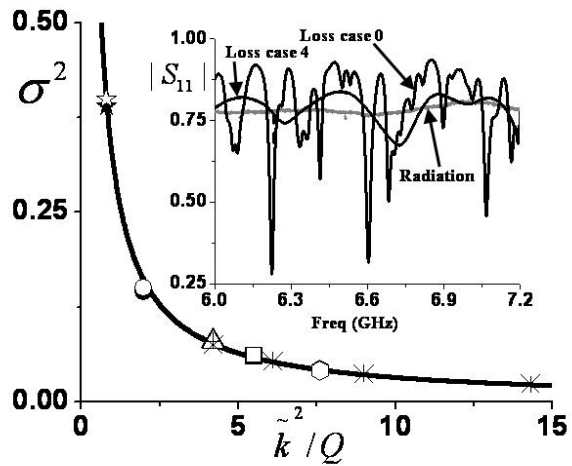

**Fig.2:**. Plot of PDF variances for $Re[z]$ (open) and $Im[z]$ (closed) for $h=7.87$ mm and 7.2-8.4 GHz versus fit parameter $\tilde{k}^2/Q$ that simultaneously fits both PDFs. Also shown are similar data for the case $h=1.78$ mm for $Re[z]$ (+) and $Im[z]$ (×) for the 6-7.2, 7.2-8.4, 9-9.75, and 11.25-12 GHz ranges. The solid line is the best fit to all the data points. Inset shows the reflection spectra $|S_{11}|$ from 6 to 7.2 GHz for two loss cases (Loss case 0: no absorber and Loss case 4: 4 strips of absorber), and the radiation case.

In Fig. 2 we test another prediction of our theory which concerns the variance of the PDFs and its relation to the control parameter $\tilde{k}^2/Q$. We can systematically change the $Q$ by changing the cavity height and the amount of microwave absorber along the interior walls, and we change $\tilde{k}^2$ by changing the frequency. Fig. 2 shows the variance of the experimental PDFs of the real and imaginary parts of $z$ compared to the $\tilde{k}^2/Q$ fit parameter for the same PDFs, for a number of cases. The open symbols (+ symbols) are the variance of the real part of $z$ for a cavity height of $h = 7.87$ mm ($h = 1.78$ mm); the closed symbols (× symbols) are the variance of the imaginary part of $z$ for $h = 7.87$ mm ($h = 1.78$ mm). The best-fit hyperbola (solid line in Fig. 2) has a coefficient of 0.32±0.01, while the theory [28] predicts it to be $1/\pi \approx 0.32$. We observe very good agreement between the theoretical prediction and our data. We also note a close overlap between the variances of Re[ $z$ ] and Im[ $z$ ] consistent with the prediction that they are equal in the limit $Q \gg 1$. This agreement is very robust experimentally and is seen independent of frequency, type and amount of loss, cavity dimensions, antenna properties, etc.

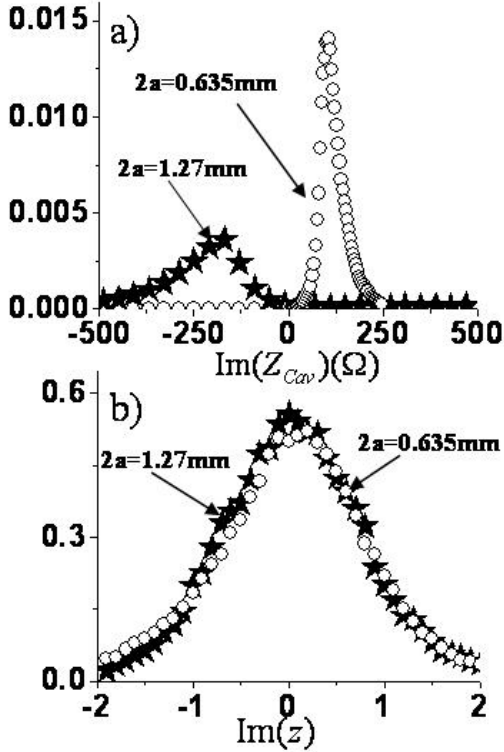

**Fig.3:** (a) shows the PDFs of the imaginary part of cavity impedance for two different antenna diameters, 2a=0.635mm (circles) and 2a=1.27mm (stars), from 9 GHz to 9.6 GHz. (b) The two curves in (a) scale together using the prescription of theory for the imaginary normalized cavity impedance.

We also test the degree of insensitivity of the universal properties of the normalized impedance PDFs to details and non-universal quantities. Working in the 9 GHz to 9.6 GHz range, we take two identical cavities and change only the diameter of the coupling wire in the antenna from $2a = 1.27$ mm to 0.635 mm. As seen in Fig. 3(a), this difference causes a dramatic change in the raw Im[$Z_{Cav}$] PDF. However, in agreement with the theoretical prediction, this difference essentially disappears in the PDFs for the appropriately scaled impedance $z$ as shown in Fig. 3(b).

The results tested here are based on very general considerations and should apply equally well to conductance measurements through quantum dots, impedance or scattering matrix measurements on electromagnetic or acoustic enclosures, and scattering experiments from nuclei and Rydberg atoms. The statistical mean and variance of PDFs can be determined *a priori* by determining the radiation impedance of the coupling geometry. The mean value of the Im[$Z_{Cav}$] distribution is equal to the imaginary part of the radiation impedance Im[$Z_{Rad}$] of the same coupling geometry. The magnitude of the reactance fluctuations corresponds to the real part of the radiation impedance (Re [$Z_{Rad}$]) in the low loss case, and then diminishes as $\sqrt{Q}$ as losses increase.

In conclusion we have examined key testable predictions for the statistics of impedance fluctuations and found satisfactory agreement on all experimental issues directly related to the theory. We find that a single parameter simultaneous fit to two independent PDFs is remarkably robust and successful, and the fit parameter is physically reasonable. The normalized cavity impedance describes universal properties of the impedance matrix fluctuations that depend only on a single control parameter.

We acknowledge useful discussions with R. Prange and S. Fishman as well as comments from Y. Fyodorov and P. Brouwer. This work was supported by the DOD MURI for the study of microwave effects under AFOSR Grant F496200110374.

---